\documentclass[conference]{IEEEtran}
\usepackage{cite}
\usepackage{graphicx} 
\usepackage{mwe}
\usepackage{tikz}
\usepackage{url}

\usetikzlibrary{shapes,arrows,positioning,arrows.meta,calc,decorations.pathreplacing,shapes,fit}
\usepackage{amsmath}
\usepackage{caption}
\usepackage{float}
\usepackage{array}
\usepackage{adjustbox}
\usepackage{subcaption}

\usepackage{capt-of,etoolbox}
\usepackage{graphicx}
\usepackage{etoolbox}
\usepackage{environ}

\newcommand{\insertfig}{\setcounter{figure}{0}\includegraphics[width=500pt]{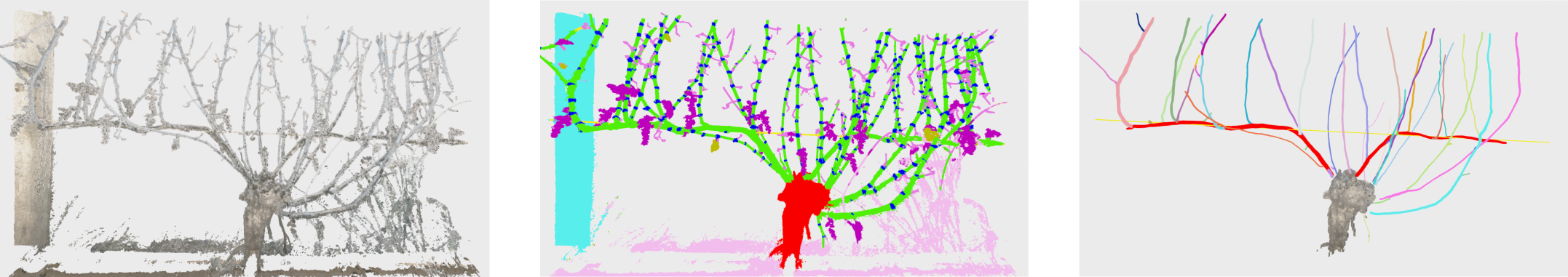}\captionof{figure}{Left to right: Input grapevine point cloud, segmented vine point cloud, and final 3D structure.}\label{fig:qualitative}}

\makeatletter

\apptocmd{\@maketitle}{\centering\insertfig}{}{}

\makeatother
\begin{document}

\title{Accurate 3D Grapevine Structure Extraction\\ from High-Resolution Point Clouds}
\author{
    \IEEEauthorblockN{Harry Dobbs\IEEEauthorrefmark{1}, Casey Peat\IEEEauthorrefmark{1}, Oliver Batchelor\IEEEauthorrefmark{2}, James Atlas\IEEEauthorrefmark{2}, Richard Green\IEEEauthorrefmark{2}}
    \IEEEauthorblockA{Department of Computer Science and Software Engineering, University of Canterbury
    \\\IEEEauthorrefmark{1}\{harry.dobbs, casey.peat\}@pg.canterbury.ac.nz
    \\\IEEEauthorrefmark{2}\{oliver.batchelor, james.atlas, richard.green\}@canterbury.ac.nz}
}

\IEEEoverridecommandlockouts
\IEEEpubid{\makebox[\columnwidth]
{979-8-3315-1877-6/24/\$31.00~\copyright2024 IEEE \hfill}} 
\maketitle
\IEEEpubidadjcol

\IEEEpeerreviewmaketitle

\begin{abstract}
Accurate 3D modelling of grapevines is crucial for precision viticulture, particularly for informed pruning decisions and automated management techniques. However, the intricate structure of grapevines poses significant challenges for traditional skeletonization algorithms. This paper presents an adaptation of the Smart-Tree algorithm for 3D grapevine modelling, addressing the unique characteristics of grapevine structures. We introduce a graph-based method for disambiguating skeletonization. Our method delineates individual cane skeletons, which are crucial for precise analysis and management. We validate our approach using annotated real-world grapevine point clouds, demonstrating an improvement of 15.8\% in the F1 score compared to the original Smart-Tree algorithm. 
This research contributes to advancing 3D grapevine modelling techniques, potentially enhancing both the sustainability and profitability of grape production through more precise and automated viticulture practices.
\end{abstract}

\section{Introduction}

Crucial to the agricultural industry, grapevines require precise management for optimal growth, health, and yield. Pruning is a critical tool for manipulating vine structure, regulating crop load, and maintaining balance \cite{christensen2000vine}. It serves three primary purposes: shaping vines to conform with trellis systems and facilitate vineyard operations; removing old wood and retaining fruiting canes or spurs for current and future seasons; and selecting fruiting wood that balances vine growth and capacity. Improper pruning can have severe consequences, including disrupted sap flows, increased disease susceptibility, yield reduction, and premature plant death \cite{gentilhomme2023towards}. Accurate 3D models of grapevines are essential for informed pruning decisions and optimised vineyard management. Precision agriculture techniques rely heavily on these models \cite{williams2023modelling}, but grapevines' intricate branching patterns and thin stems present significant challenges for traditional skeletonization algorithms.

\subsection{Motivation}
While our previously introduced Smart-Tree \cite{dobbs2023smart} point cloud skeletonization algorithm has shown promise for various tree species, its direct application to grapevines is suboptimal. The dense, intricate structure of grapevines, with numerous canes growing nearby and often intertwining, poses unique challenges. This complexity can lead to frequent failure cases (detailed in Section \ref{sec:failure_cases}) where multiple canes are incorrectly merged or branching topology is misinterpreted, significantly impacting subsequent analyses.

This paper presents a comprehensive study on adapting Smart-Tree for 3D grapevine modelling. We discuss specific algorithm modifications, including integrating prior knowledge and refining processing techniques. We validate our adapted approach using annotated real-world grapevine point cloud data, demonstrating improved accuracy and robustness in handling grapevine structures.

The main contributions of this paper are:
\begin{enumerate}
\item Adaptation of the Smart-Tree algorithm to grapevine-specific characteristics and requirements.
\item Incorporation of domain-specific knowledge to enhance accuracy and robustness.
\item Development of methods to delineate individual cane skeletons for precise analysis and management.
\item Validation of the adapted approach using real-world grapevine point cloud data.
\end{enumerate}

This research advances 3D grapevine modelling, enabling precise viticulture techniques. Our tailored skeletonization method contributes to innovative vineyard management tools, aiming to enhance grape production's sustainability and profitability.
\newpage


    
    
    
    


\section{Related Work}
Recent advancements in agricultural robotics have focused on automating tasks such as harvesting, yield estimation, and pruning. This section reviews relevant work in vineyard automation and 3D plant modelling.
\subsection{Agricultural Robotics and Image Analysis}
Robotic systems for fruit harvesting have been developed for crops like apples, and kiwifruit \cite{silwal2017design, williams2019robotic}, demonstrating the potential for agricultural automation. Yield estimation techniques have progressed from seasonal model-based predictions \cite{basso2019seasonal} to direct fruit counting using deep learning \cite{koirala2019deep}.

\subsection{Grapevine modelling and Pruning}
Several approaches have been proposed for grapevine modelling and automated pruning:

\begin{itemize}
    \item Botterill et al. \cite{botterill2017robot} developed a pruning robot using a trinocular camera system for 3D reconstruction but faced challenges with occlusion and lighting control.
    
    \item Gentilhomme et al. \cite{gentilhomme2023towards} used an hourglass CNN to identify nodes and reveal skeletal structure but struggled with occlusions and erroneous connections.
    
    \item Silwal et al. \cite{silwal2022bumblebee} proposed the Bumblebee system for spur pruning, achieving 87\% pruning accuracy, but with limited evaluation of the 3D modelling performance.
    
    \item Williams et al. \cite{williams2023modelling} presented a vision system for autonomous cane pruning, generating 3D models with skeletonized cane structures. Their system modelled 51.45\% of canes entirely, with an additional 35.51\% 
\end{itemize}

\subsection{Limitations in Current Approaches}
Despite these advancements, most existing approaches rely on 2D image processing, limited 3D reconstruction from stereo vision, or simplified 3D models. Using high-resolution point clouds for detailed grapevine structure modelling remains largely unexplored. This gap is significant as high-resolution point cloud models can provide more accurate and comprehensive representations of grapevine structures, potentially improving performance in pruning, yield estimation, and overall vine management tasks.

Our work addresses this gap by adapting the Smart-Tree algorithm to work with high-resolution 3D point cloud data of grapevines, aiming to advance precision viticulture techniques.

\section{Challenges in Grapevine Skeletonization}
\label{sec:failure_cases}
Skeletonization algorithms face significant challenges when applied to grapevines' complex structures. We identify three primary failure cases: broken canes, cane jumping, and incorrect cane correspondence. These issues are particularly prevalent in methods utilizing shortest-path algorithms, which are prevalent in the field.

\subsection{Cane Jumping}
\label{sec:challenges_cane_jumping}
Cane jumping is widespread in skeletonization algorithms that rely on shortest-path computations. This phenomenon occurs when the algorithm incorrectly "jumps" from one cane to another where they grow close together or cross. The cane jumping error typically occurs when the algorithm correctly traces the first cane (C1) from its base to its tip. Still, when tracing the second cane (C2), continuing along C1 provides a shorter route to the termination points. As a result, C2 erroneously terminates where it meets C1, and the algorithm generates an additional, incorrect cane (C3) to account for the remaining untraced points. This error leads to prematurely terminated canes and the generation of erroneous additional canes, significantly distorting the vine's structure (Fig. \ref{fig:cane_jumping}).

\definecolor{trunkbrown}{RGB}{120,68,33}
\definecolor{tikzfillthree}{RGB}{61,166,0}
\definecolor{tikzfillfour}{RGB}{255,0,255}

\tikzset{
  branch/.style={line width=1.5mm, color=trunkbrown},
  cane/.style={{Triangle[length=3mm,width=4mm]}-, line width=0.8mm},
  label/.style={font=\fontfamily{phv}\selectfont\bfseries}
}

\newcommand{\basegrapevine}[2]{
  \begin{scope}[shift={(#1,#2)}]
    \fill[fill=trunkbrown] (0,0.5) rectangle (0.5,1.5);
    \draw[branch] (0.3,1.4) .. controls (1,1.7) and (2.6,1.2) .. (2.7,1.7);
  \end{scope}
}


\newcommand{\caneone}{
  \draw[cane, red] (0.2,1.5) -- (2,3.6) node[right,label] {C1};
  \fill[red] (2,3.6) circle (3pt);
}

\newcommand{\caneoneB}{
  \draw[cane, red] (0.2,1.5) -- (1.1,2.6) node[left=0.25cm,label] {C1};
  \fill[red] (1.1,2.6) circle (3pt);
}

\newcommand{\canetwoA}{
  \draw[cane, blue] (1.1,2.6) -- (0.1,3.6) node[left, label] {C2};
  \fill[blue] (0.1,3.6) circle (3pt);
}

\newcommand{\canetwoB}{
  \draw[cane, blue] (2,1.5) -- (0.0,3.5) node[left, label] {C2};
  \fill[blue] (0.0,3.5) circle (3pt);
}

\newcommand{\canethree}{
  \draw[cane, tikzfillthree] (2,1.5) -- (1.1,2.6) node[right=0.25cm,label, text=tikzfillthree] {C3};
  \fill[tikzfillthree] (1.1,2.6) circle (3pt);
}

\newcommand{\canefour}{
  \draw[cane, tikzfillfour] (1.1,2.6)  -- (2,3.6) node[right=0.25cm,label, text=tikzfillfour] {C4};
  \fill[tikzfillfour] (2,3.6) circle (3pt);
}

\newcommand{\Agrapevine}[2]{
  \basegrapevine{#1}{#2}
  \begin{scope}[shift={(#1,#2)}]
    \caneone
    \canethree
    \canetwoA
  \end{scope}
}

\newcommand{\Bgrapevine}[2]{
  \basegrapevine{#1}{#2}
  \begin{scope}[shift={(#1,#2)}]
    \caneone
    \canetwoB
  \end{scope}
}

\newcommand{\splitgrapevine}[2]{
  \basegrapevine{#1}{#2}
  \begin{scope}[shift={(#1,#2)}]
    \caneoneB
    \canetwoA
    \canethree
    \canefour
  \end{scope}
}

\newcommand{\connectiongrapevine}[2]{
  \basegrapevine{#1}{#2}
  \begin{scope}[shift={(#1,#2)}]
    \caneoneB
    \canetwoA
    \canethree
    \canefour

    \draw[loosely dashed, red, line width=1.5pt]  (1.1,2.6) circle (0.3cm);
    \draw[loosely dashed, blue, line width=1.5pt]  (0.1,3.6) circle (0.3cm);
    \begin{scope}[rotate around={36:(1.1,2.6)}]
        \draw[loosely dashed, tikzfillthree, line width=1.5pt] (1.1,2.6) circle (0.3cm);
    \end{scope}
    \draw[loosely dashed, tikzfillfour, line width=1.5pt] (2,3.6) circle (0.3cm);

  \end{scope}
}

\newcommand{\Cgrapevine}[2]{
  \Agrapevine{#1}{#2}
}

\newcommand{\Dgrapevine}[2]{
  \Bgrapevine{#1}{#2}
}

\newcommand{\Egrapevine}[2]{
  \splitgrapevine{#1}{#2}
}


\begin{figure}[htbp]
  \centering
  \begin{tikzpicture}[scale=0.8]
    \Cgrapevine{0}{0}
    \Dgrapevine{0.5\linewidth}{0}
  \end{tikzpicture}
  \caption{Cane Jumping: Left: Incorrect output showing C2 terminating at C1 and formation of erroneous C3. Right: Target output with correctly separated canes.}
  \label{fig:cane_jumping}
\end{figure}
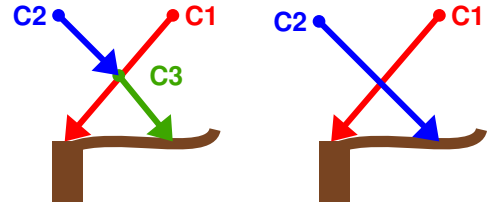



\subsection{Incorrect Cane Correspondence}
\label{sec:challenges_cane_correspodence}
Incorrect cane correspondence occurs when the algorithm erroneously merges multiple distinct canes into a single skeletal structure. This issue arises when tracing a path that inadvertently incorporates portions of numerous canes. Fig. \ref{fig:incorrect_correspondence} illustrates this problem: the left image shows two distinct canes incorrectly labelled as C1 (red and blue) merged at their intersection. This results in merged canes that should be separate, prematurely terminated canes where they intersect with previously traced paths and a misrepresentation of the vine's true topology. The right image displays the correct output, where the red cane remains C1, and the blue cane is properly identified as C2.


\newcommand{\jumpcaneoneA}{
  \draw[cane, red] (2.0,1.4) -- (1.05,2.45) .. controls (1.0,2.5) and (1.0,2.55) .. (1.05,2.6) -- (2.0,3.5) node[right, label] {C1};
  \fill[red] (2.0,3.5) circle (3pt);
  \fill[red] (2.0,3.5) circle (3pt);
}

\newcommand{\jumpcanetwoA}{
  \draw[cane, blue]  (1.0, 2.55) --  (0.0, 3.6) node[left, label] {C2};
  \fill[blue] (0.0, 3.6) circle (3pt);
}

\newcommand{\jumpcaneoneB}{
  \draw[cane, red] (1.0,2.55) -- (2.0,3.5)node[right, label] {C1};
  \fill[red] (2.0,3.5) circle (3pt);}

\newcommand{\jumpcanetwoB}{
  \draw[cane, blue] (2.0,1.4) -- (0.0, 3.6) node[left, label] {C2};
  \fill[blue] (0.0, 3.6) circle (3pt);
}

\newcommand{\jumpAgrapevine}[2]{
  \basegrapevine{#1}{#2}
  \begin{scope}[shift={(#1,#2)}]
    \jumpcaneoneA
    \jumpcanetwoA
  \end{scope}
}

\newcommand{\jumpBgrapevine}[2]{
  \basegrapevine{#1}{#2}
  \begin{scope}[shift={(#1,#2)}]
    \jumpcaneoneB
    \jumpcanetwoB
  \end{scope}
}

\newcommand{\jumpCgrapevine}[2]{
  \jumpAgrapevine{#1}{#2}
}

\newcommand{\jumpDgrapevine}[2]{
  \jumpBgrapevine{#1}{#2}
}


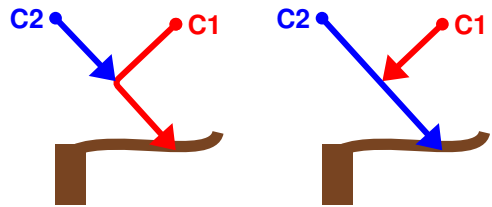
\begin{figure}[htbp]
  \centering
  \begin{tikzpicture}[scale=0.8]
    \jumpCgrapevine{0}{0}
    \jumpDgrapevine{0.5\linewidth}{0}
  \end{tikzpicture}
\caption{Cane Correspondence: Left: Incorrect output showing merged canes both labelled C1. Right: Correct output with properly separated canes C1 and C2.}
\label{fig:incorrect_correspondence}
\end{figure}


\subsection{Broken Canes}
\label{sec:challenges_broken_canes}
Broken canes stem from gaps in point cloud data or segmentation inaccuracies, resulting in disconnected components in the initial data. The main challenge lies in correctly identifying and connecting these components. Failure to do so leads to underestimated cane lengths and incorrect vine topology, potentially affecting critical decisions in vineyard management.

\section{Method}
\subsection{Overview}
Our grapevine skeletonization method comprises eight steps. The process begins with point cloud generation from high-resolution images (4000 x 3000 pixels) captured by an autonomous RTK robot using a rig of 6 cameras with 14cm baseline spacing and high-powered LED lights. The images from each side of the vine undergo initial alignment through structure-from-motion using Agisoft Metashape \cite{agisoft2024metashape}, followed by NeRF \cite{wang2021nerf} based reconstruction to generate point clouds. The side views are then aligned using ZeroNeRF \cite{peat2022zero} to create a single sub-millimetre resolution point cloud. This is followed by segmentation using a U-Net architecture implemented as a Sparse Convolutional Neural Network. We then perform wire and trunk identification using RANSAC and neighbourhood graph techniques. The initial skeletonization uses the Smart-Tree algorithm \cite{dobbs2023smart}, followed by a broken cane repair step to address disconnected components. We apply an iterative cane fitting process at multiple stages: on the initial skeleton from Smart-Tree, when canes are broken into segments during refinement, and on the final skeleton. This fitting process refines skeleton accuracy by optimizing point-to-cane distances using a Welsch loss function and incorporating smoothness regularization. Skeleton refinement is achieved through a novel approach that breaks canes into segments and reconstructs them based on a smoothness heuristic. Finally, we perform node detection using DBSCAN clustering and extract DOLPHIN metrics for pruning decisions. This comprehensive approach enables us to generate accurate skeletal representations of complex grapevine structures, addressing challenges such as occlusions and segmentation errors. Our method is developed and evaluated using a dataset of 55 vines, partitioned into training, validation, and testing subsets of 30, 10, and 15 vines, respectively. To establish ground truth, we use a custom semi-automatic cylinder labelling technique for annotating skeletons on the point clouds, ensuring accurate reference data for training and evaluation.

\subsection{Segmentation}
Our method begins by segmenting the vineyard point cloud into its components, classifying each point into one of several classes: trunk, canes, nodes, wires, posts, tendrils, rachis, ground plane, and leaves. We implement a U-Net architecture with residual connections as a Sparse Convolutional Neural Network (SCNN) using SpConv \cite{spconv2022} and PyTorch \cite{paszke2019pytorch}. Points are voxelized at a resolution of 5mm before being processed by the network. The model is trained on sparsely labelled real-world data, capturing the complexity of actual vineyard structures while minimizing labelling effort. 
 Our segmentation model achieves high performance across all classes, with class-specific F1-scores ranging from 83.3\% for nodes to 99.6\% for posts. For a detailed description of our segmentation approach, including architecture, training strategies, and performance evaluation, we refer readers to our previous paper \cite{dobbs2024integrating}.

\subsection{Wire and Trunk Identification}
\textbf{Wire Identification:} Wire identification is performed through an iterative process using RANSAC \cite{fischler1981random} based 3D line fitting. We implement this using our PyTorch-based CUDA RANSAC library \cite{Dobbs_torch_ransac3d} for efficient parallel computation. The algorithm filters wire-class points from the point cloud and applies a two-pass RANSAC approach with line distance thresholds of 0.02 and 0.008, respectively. Identified wires are modelled as start and end points with a radius of 0.002 units. The process continues until fewer than 15,000 wire points remain or six wires are detected.

\textbf{Trunk Identification:} Trunk identification establishes root points for skeletal graph generation. We create a neighbourhood graph by connecting each predicted trunk point to its neighbours within a 0.02m radius. This graph's most significant connected component is selected as the set of trunk points, effectively filtering out erroneous segmentation results. Due to the non-convex nature of grapevine trunks, we represent them as a voxel-grid. This approach comprehensively represents and defines a keep-out zone for the robotic pruning arm.

\subsection{Skeletonization}
We use the Smart-Tree algorithm \cite{dobbs2023smart}, which uses a sparse convolutional neural network to predict the radius and direction to the medial axis for each point in the vine point cloud. The algorithm constructs a constrained neighbourhood graph and extracts the skeleton through path-finding operations. We encourage the reader to read the original paper \cite{dobbs2023smart} for complete details. Due to occlusions, gaps in the point cloud, or segmentation errors, this initial process may produce multiple disconnected skeletal components rather than a single, fully connected skeleton. 

\subsection{Broken Cane Repair}
We implement a repair algorithm to address disconnected skeletal components that evaluate potential connections based on distance and angle criteria. For a connected branch endpoint $\mathbf{B}$ and disconnected component base point $\mathbf{d}$, the algorithm:

\begin{enumerate}
    \item Computes direction vectors $\vec{v}_1$ (along connected branch), $\vec{v}_2$ (along disconnected component), and $\vec{v}_3$ (from $\mathbf{B}$ to $\mathbf{d}$)
    \item Calculates angular differences between the directions of: $\theta_1$ ($\vec{v}_1$ and $\vec{v}_3$), $\theta_2$ ($\vec{v}_2$ and $\vec{v}_3$), and $\theta_3$ ($\vec{v}_1$ and $\vec{v}_2$)
    \item Measures distance $d$ between $\mathbf{B}$ and $\mathbf{d}$
    \item Connects branches if $d \leq 0.15$ m and all $\theta_i \leq 45^\circ$
\end{enumerate}

This process iterates until no further valid connections can be made, resulting in a more coherent skeletal representation that maintains a realistic branching structure.

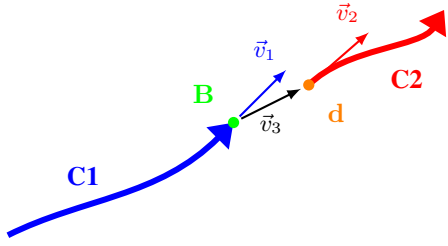
\begin{figure}[htbp]
    \centering
    \begin{tikzpicture}[
        every node/.style={font=\sffamily},
        cane/.style={line width=0.8mm, -{Triangle[length=3mm,width=4mm]}},
        point/.style={circle, inner sep=1.5pt},
        vector/.style={->, >=latex, thick},
        label/.style={font=\small},
        pointlabel/.style={font=\bfseries},
        scale=1.0
    ]
    \draw[cane, blue] (0,0) .. controls (1,0.5) and (2,0.5) .. (3,1.5) node[point, fill=green] (b) {} node[above, font=\bfseries, xshift=-2cm, yshift=-1.0cm] {C1};

    \draw[cane, red] (4,2) node[point, fill=orange] (d) {} .. controls (4.5,2.5) and (5.5,2.5) .. (5.8,3) node[above, font=\bfseries, xshift=-0.5cm, yshift=-1.2cm] {C2};
    
    \node[pointlabel, above left=2pt of b, text=green] {$\mathbf{B}$};
    \node[pointlabel, below right=2pt of d, text=orange] {$\mathbf{d}$};
    
    \draw[vector, blue] (b) -- ++ (0.7,0.7) node[above left, label] {$\vec{v}_1$};
    
    \draw[vector, red] (d) -- ++(0.8,0.7) node[above left, label] {$\vec{v}_2$};
    
    \draw[vector, black] (b) -- (d) node[midway, below, label] {$\vec{v}_3$};
    
    
\end{tikzpicture}
\caption{Illustration of the Broken Cane Repair algorithm showing the connected cane (C1), disconnected cane (C2), key points ($\mathbf{B}$ and $\mathbf{d}$), vectors ($\vec{v}_1$, $\vec{v}_2$, $\vec{v}_3$), and angles ($\theta_1$, $\theta_2$, $\theta_3$) used for evaluation.}
\label{fig:broken-cane-repair}
\end{figure}

\subsection{Skeleton Refinement}
We introduce a refinement step to address the challenges mentioned in Section \ref{sec:challenges_cane_jumping} and Section \ref{sec:challenges_cane_correspodence}.
The high-level explanation of this method is to break the canes into smaller segments and then identify which segments belong to the same cane based on a smoothness heuristic.
\newcommand{\tikzscale}{1}
\newcommand{\tikznodesize}{0.7cm}
\newcommand{\tikzfontsize}{7pt}
\newcommand{\tikzarrowwidth}{2pt}
\newcommand{\tikzbaseoffset}{0.8}
\newcommand{\tikzdoubleoffset}{1.6}
\newcommand{\tikzquarteroffset}{0.4}
\newcommand{\tikzfont}{\bfseries\sffamily}
\newcommand{\tikzfontcolour}{white}

\newcommand{\darkbrown}{rgb,255:red,120; green,68; blue,33}
\newcommand{\tikzfillone}{rgb,255:red,255; green,0; blue,0}
\newcommand{\tikzfilltwo}{rgb,255:red,0; green,0; blue,255}
\newcommand{\tikzfillthree}{rgb,255:red,61; green,166; blue,0}
\newcommand{\tikzfillfour}{rgb,255:red,255; green,0; blue,255}
\newcommand{\verticalspace}{10.0pt}
\newcommand{\imageheight}{2.89cm}
\newcommand{\vinescale}{0.85}

\newcommand{\subfigurewidth}{0.48\linewidth}

\begin{figure}[htbp]
    \centering
    \begin{subfigure}[b]{\subfigurewidth}
        \centering
        \begin{tikzpicture}[scale=\vinescale]
            \Agrapevine{0}{0}
        \end{tikzpicture}
        \caption{Initial skeleton}

    \end{subfigure}
    \hfill
    \begin{subfigure}[b]{\subfigurewidth}
        \centering
        \adjustbox{valign=c}{
        \begin{tikzpicture}[
            scale=\tikzscale,
            every node/.style={circle, draw, minimum size=\tikznodesize, font={\tikzfont\fontsize{\tikzfontsize}{\tikzfontsize}\selectfont}, text=\tikzfontcolour},
            arrow/.style={->, >=stealth, line width=\tikzarrowwidth}
        ]
        \node[fill=\darkbrown] (root) at (0,0) {};
        \node[fill=\tikzfillone] (C1) at (-\tikzbaseoffset,\tikzbaseoffset) {C1};
        \node[fill=\tikzfilltwo] (C2) at (-\tikzdoubleoffset,\tikzdoubleoffset) {C2};
        \node[fill=\tikzfillthree] (C3) at (\tikzbaseoffset,\tikzbaseoffset) {C3};
        \draw[arrow] (C2) -- (C1);
        \draw[arrow] (C1) -- (root);
        \draw[arrow] (C3) -- (root);
        \end{tikzpicture}
        }
        \caption{Graph representation}
    \end{subfigure}
    
    \vspace{\verticalspace}
    
    \begin{subfigure}[b]{\subfigurewidth}
        \centering
        \begin{tikzpicture}[scale=\vinescale]
            \Egrapevine{0}{0}
        \end{tikzpicture}    
        \caption{Split canes}
    \end{subfigure}
    \hfill
    \begin{subfigure}[b]{\subfigurewidth}
        \centering
        \adjustbox{valign=c}{
        \begin{tikzpicture}[
            scale=\tikzscale,
            every node/.style={circle, draw, minimum size=\tikznodesize, font={\tikzfont\fontsize{\tikzfontsize}{\tikzfontsize}\selectfont}, text=\tikzfontcolour},
            arrow/.style={->, >=stealth, line width=\tikzarrowwidth}
        ]
        \node[fill=\darkbrown] (root) at (0,0) {};
        \node[fill=\tikzfillone] (C1) at (-\tikzbaseoffset,\tikzbaseoffset) {C1};
        \node[fill=\tikzfilltwo] (C2) at (-\tikzdoubleoffset,\tikzdoubleoffset) {C2};
        \node[fill=\tikzfillthree] (C3) at (\tikzbaseoffset,\tikzbaseoffset) {C3};
        \node[fill=\tikzfillfour] (C4) at (0,\tikzdoubleoffset) {C4};
        \draw[arrow] (C2) -- (C1);
        \draw[arrow] (C1) -- (root);
        \draw[arrow] (C4) -- (C1);
        \draw[arrow] (C3) -- (root);
        \end{tikzpicture}
        }
        \caption{Updated graph}
    \end{subfigure}
    
    \vspace{\verticalspace}
    
    \begin{subfigure}[b]{\subfigurewidth}
        \centering
        \begin{tikzpicture}[scale=\vinescale]
            \connectiongrapevine{0}{0}
        \end{tikzpicture}            
        \caption{Potential connection search}
    \end{subfigure}
    \hfill
    \begin{subfigure}[b]{\subfigurewidth}
        \centering
        \adjustbox{valign=c}{
        \begin{tikzpicture}[
            scale=\tikzscale,
            every node/.style={circle, draw, minimum size=\tikznodesize, font={\tikzfont\fontsize{\tikzfontsize}{\tikzfontsize}\selectfont}, text=\tikzfontcolour},
            arrow/.style={->, >=stealth, line width=\tikzarrowwidth},
            dashedarrow/.style={->, >=stealth, dashed, line width=\tikzarrowwidth}
        ]
        \node[fill=\darkbrown] (root) at (0,0) {};
        \node[fill=\tikzfillone] (C1) at (-\tikzbaseoffset,\tikzbaseoffset) {C1};
        \node[fill=\tikzfilltwo] (C2) at (-\tikzdoubleoffset,\tikzdoubleoffset) {C2};
        \node[fill=\tikzfillthree] (C3) at (\tikzbaseoffset,\tikzbaseoffset) {C3};
        \node[fill=\tikzfillfour] (C4) at (0,\tikzdoubleoffset) {C4};
        \draw[arrow] (C2) -- (C1);
        \draw[arrow] (C1) -- (root);
        \draw[arrow] (C4) -- (C1);
        \draw[arrow] (C3) -- (root);
        \draw[dashedarrow, out=30, in=100, looseness=1.8] (C2) to (C3);
        \draw[dashedarrow] (C4) to (C3);
        \end{tikzpicture}
        }
        \caption{Graph with potential links}
    \end{subfigure}
    
    \vspace{\verticalspace}
    
    \begin{subfigure}[b]{\subfigurewidth}
        \centering
        \adjustbox{valign=c}{
        \begin{tikzpicture}[
            scale=\tikzscale,
            every node/.style={circle, draw, minimum size=\tikznodesize, font={\tikzfont\fontsize{\tikzfontsize}{\tikzfontsize}\selectfont}, text=\tikzfontcolour},
            arrow/.style={->, >=stealth, line width=\tikzarrowwidth},
            dashedarrow/.style={->, >=stealth, dashed, line width=\tikzarrowwidth}
        ]
        \node[fill=\darkbrown] (root) at (0,0) {};
        \node[fill=\tikzfillone] (C1) at (-\tikzbaseoffset,\tikzbaseoffset) {C1};
        \node[fill=\tikzfilltwo] (C2) at (-\tikzdoubleoffset,\tikzdoubleoffset) {C2};
        \node[fill=\tikzfillthree] (C3) at (\tikzbaseoffset,\tikzbaseoffset) {C3};
        \node[fill=\tikzfillfour] (C4) at (0,\tikzdoubleoffset) {C4};
        \draw[arrow, red] (C2) -- (C1);
        \draw[arrow, orange] (C1) -- (root);
        \draw[arrow, green] (C4) -- (C1);
        \draw[arrow, orange] (C3) -- (root);
        \draw[dashedarrow, green, out=30, in=100, looseness=1.8] (C2) to (C3);
        \draw[dashedarrow, red] (C4) to (C3);
        \end{tikzpicture}
        }
        \caption{Weighted connections}
    \end{subfigure}
    \hfill
    \begin{subfigure}[b]{\subfigurewidth}
        \centering
        \adjustbox{valign=c}{
        \begin{tikzpicture}[
            scale=\tikzscale,
            every node/.style={circle, draw, minimum size=\tikznodesize, font={\tikzfont\fontsize{\tikzfontsize}{\tikzfontsize}\selectfont}, text=\tikzfontcolour},
            arrow/.style={->, >=stealth, line width=\tikzarrowwidth}
        ]
        \node[fill=\darkbrown] (root) at (0,0) {};
        \node[fill=\tikzfillone] (C1) at (-\tikzbaseoffset,\tikzbaseoffset) {C1};
        \node[fill=\tikzfillfour] (C4) at (-\tikzdoubleoffset,\tikzdoubleoffset) {C4};
        \node[fill=\tikzfillthree] (C3) at (\tikzbaseoffset,\tikzbaseoffset) {C3};
        \node[fill=\tikzfilltwo] (C2) at (\tikzdoubleoffset,\tikzdoubleoffset) {C2};
        \draw[green, arrow] (C2) -- (C3);
        \draw[orange, arrow] (C1) -- (root);
        \draw[green, arrow] (C4) -- (C1);
        \draw[orange, arrow] (C3) -- (root);
        \end{tikzpicture}
        }
        \caption{Minimum spanning tree}
    \end{subfigure}
    
    \vspace{\verticalspace}
    
    \begin{subfigure}[b]{\subfigurewidth}
        \centering
        \begin{tikzpicture}[scale=\vinescale]
            \Bgrapevine{0}{0}
        \end{tikzpicture}   
        \caption{Final skeleton}
    \end{subfigure}
    \hfill
    \begin{subfigure}[b]{\subfigurewidth}
        \centering
        \adjustbox{valign=c}{
        \begin{tikzpicture}[
            scale=\tikzscale,
            every node/.style={circle, draw, minimum size=\tikznodesize, font={\tikzfont\fontsize{\tikzfontsize}{\tikzfontsize}\selectfont}, text=\tikzfontcolour},
            arrow/.style={->, >=stealth, line width=\tikzarrowwidth}
        ]
        \node[fill=\darkbrown] (root) at (0,0) {};
        \node[fill=\tikzfillone] (C1) at (-\tikzbaseoffset,\tikzbaseoffset) {C1};
        \node[fill=\tikzfilltwo] (C2) at (\tikzbaseoffset,\tikzbaseoffset) {C2};
        \draw[orange, arrow] (C1) -- (root);
        \draw[orange, arrow] (C2) -- (root);
        \end{tikzpicture}
        }
        \caption{Final graph representation}
    \end{subfigure}
    
    
    \caption{Skeleton refinement process and corresponding graph representations for grapevine cane delineation (illustrated in 2D)}
    \label{fig:skeletonrefinement}
\end{figure}
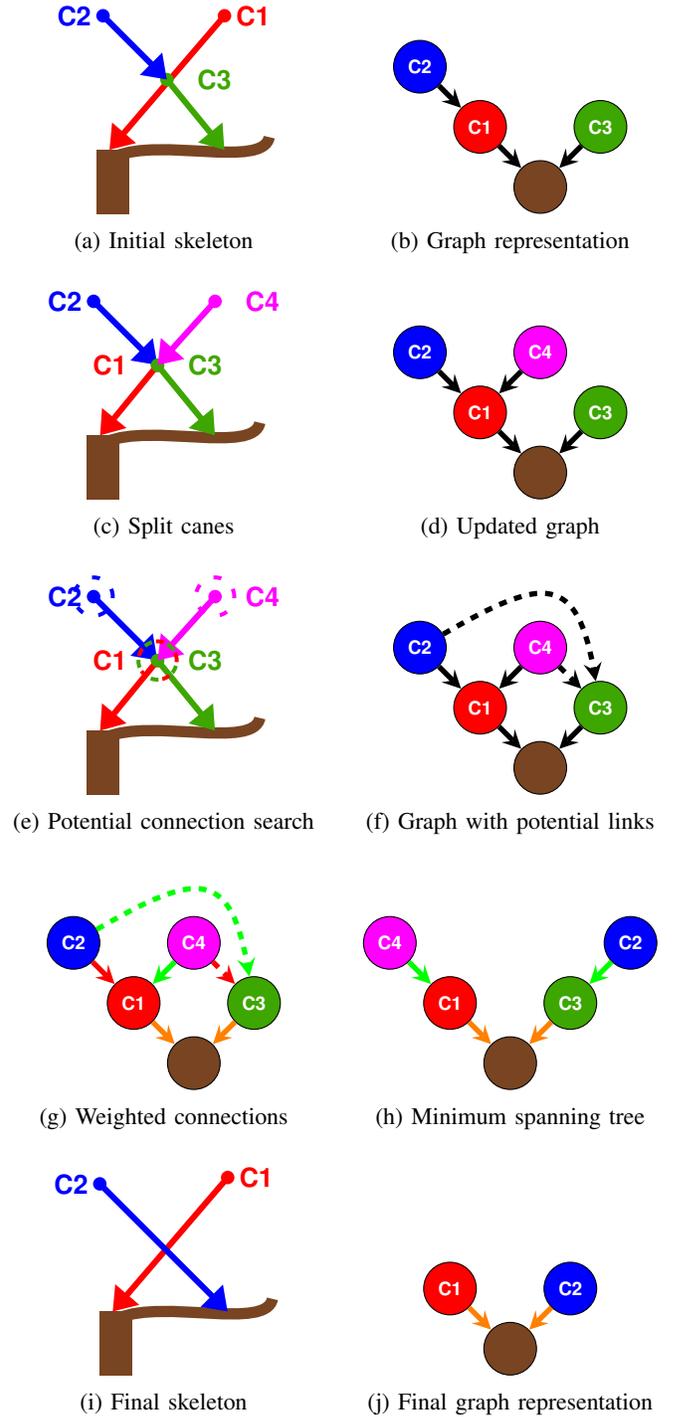

Our method for reconstructing accurate vine skeletons consists of the following steps (illustrated in Fig. \ref{fig:skeletonrefinement}):
\begin{enumerate}
\item We begin with an initial, incorrectly reconstructed skeleton as described in Section \ref{sec:challenges_cane_jumping}. Each node in the nodal representation corresponds to a cane in the skeleton (Fig. \ref{fig:skeletonrefinement}a,b).
\item When a cane intersects with another (parent) cane, we split the parent cane into multiple canes. For example, C2 intersecting C1 results in C1 splitting into C1 and C4 (Fig. \ref{fig:skeletonrefinement}c,d).

\item The end of each cane then searches within a defined radius to identify potential connections to other canes. These potential connections are represented as dashed edges in the nodal representation (Fig. \ref{fig:skeletonrefinement}e,f).

\item We compute the likelihood of these cane connections based on a smoothness heuristic. This heuristic combines two main components:
\begin{itemize}
    \item Direction alignment: Evaluates how well the directions of two potentially connected cane segments align. It considers the parent branch's last segment and the child branch's first segment.
    \item Radius difference: Compares the radii of the connecting ends of two cane segments.
\end{itemize}
The combined smoothness heuristic $H$ for connecting two cane segments is given by:
\begin{equation}
    H = w_a \cdot A + w_r \cdot R
\end{equation}
where $A$ is the alignment score, $R$ is the radius difference score, and $w_a$ and $w_r$ are their respective weights (default: $w_a = 0.8$, $w_r = 0.2$).

Edges are colour-coded: green edges indicate a higher likelihood of connection (lower weight), while red edges suggest a lower likelihood (higher weight) due to less smooth transitions between canes (Fig. \ref{fig:skeletonrefinement}g).

\item Using these computed weights, we extract the Minimum Spanning Tree (MST) from the graph (Fig. \ref{fig:skeletonrefinement}h).

\item Finally, we merge canes connected by edges with weights below a certain threshold, as these likely belong to the same cane. This process produces the desired output skeleton (Fig. \ref{fig:skeletonrefinement}i,j).
\end{enumerate}
The merging process is implemented using a post-order traversal of the MST. This algorithm iteratively merges branches with the lowest weighted child below a threshold, refining the skeleton structure to more accurately represent the true vine architecture, as seen in the progression from Fig. \ref{fig:skeletonrefinement}a to \ref{fig:skeletonrefinement}j.

\subsubsection{Extraction of Bearer Canes}
We focus on the leftmost and rightmost canes to identify bearer canes, analyzing their proximity to the trellis wire. We track each potential bearer cane from the vine base upwards, splitting it at a predefined distance threshold above the wire. The lower segment is classified as the bearer cane, while the upper segment becomes a standard cane. 

\subsection{Cane Fitting} 
To refine skeleton accuracy, we implement an iterative cane fitting process at multiple stages: on the initial Smart-Tree skeleton, during segment-wise refinement, and on the final skeleton. For each cane or segment, we calculate point-to-cane distances, apply a Welsch loss function $L_\text{welsch}(x, \mu, \sigma) = 1 - \exp(-\frac{1}{2}(\frac{x - \mu}{\sigma})^2)$ for robustness against outliers, and compute a total loss $L_\text{total} = \sum_{i} (w_i \cdot L_i) + L_\text{fit}$ incorporating regularization terms. These terms include cane smoothness to prevent abrupt changes in direction and radius smoothness to ensure gradual thickness transitions. Cane parameters are updated via gradient descent, and the process repeats until convergence or an iteration limit is reached. For computational efficiency, we employ random point subset sampling during loss calculation. 

\subsection{Node Detection}
After point cloud segmentation, node detection is performed in two stages: spatial clustering and cane association. Node-labeled points are clustered using DBSCAN (eps=0.02m, min samples=10) to identify distinct nodes, with each node's position computed as the mean of its cluster points. Nodes are associated with their nearest vineyard cane.

\section{Results}
\subsection{Evaluation Metric for Vine Skeletonization}
We propose a comprehensive evaluation metric to assess the accuracy of vine skeleton extraction from point cloud data. This metric compares predicted vine skeleton paths with ground truth paths, each represented by a sequence of 3D coordinates and associated radii.

\subsubsection{Path Interpolation and IoU Calculation}
For each pair of vine paths $(g_i, p_j)$, where $g_i$ is a ground truth path and $p_j$ is a predicted path, we:
\begin{enumerate}
    \item Interpolate both paths at fixed intervals $\delta$ (default 1mm) to account for varying point densities.
    \item Calculate the Intersection over Union (IoU), considering the volumetric overlap of vine branches:
    \[
    IoU(g_i, p_j) = \frac{|I(g_i, p_j)|}{|U(g_i, p_j)|}
    \]
    Where $I(g_i, p_j)$ represents the intersection of the paths within their respective radii, and $U(g_i, p_j)$ is their union.
\end{enumerate}

\subsubsection{Optimal Path Matching}
We use the linear sum assignment problem solver from SciPy \cite{2020SciPy-NMeth} to find the optimal matching between ground truth and predicted vine paths. This implementation is based on a modified Jonker-Volgenant algorithm \cite{crouse2016implementing}. We construct a cost matrix $C$ where:
\[
C_{ij} = 1 - IoU(g_i, p_j)
\]
This approach ensures optimal one-to-one matching, handling cases where the number of ground truth paths differs from the number of predicted paths.

\subsubsection{Threshold-based Evaluation and Metrics}
Using an IoU threshold $\tau$ (default 0.75), we define the set of correct matches $M$ and compute the following metrics:
\begin{itemize}
    \item Correctly Identified Canes: $|M|$
    \item Undetected Canes: $|G| - |M|$, where $G$ is the set of ground truth paths
    \item Erroneously Detected Canes: $|P| - |M|$, where $P$ is the set of predicted paths
    \item Precision: $Precision = \frac{|M|}{|P|}$
    \item Recall: $Recall = \frac{|M|}{|G|}$
    \item F1-Score: $F1 = 2 \cdot \frac{Precision \cdot Recall}{Precision + Recall}$
\end{itemize}

\subsubsection{Weighted Metrics}
To account for the varying importance of canes based on their length, we introduce weighted versions of our metrics:
\begin{itemize}
    \item Weighted IoU: $wIoU(g_i, p_j) = IoU(g_i, p_j) \cdot \frac{length(g_i) + length(p_j)}{2}$
    \item Weighted Precision: $wPrecision = \frac{\sum_{(i,j) \in M} length(p_j)}{\sum_{j \in P} length(p_j)}$
    \item Weighted Recall: $wRecall = \frac{\sum_{(i,j) \in M} length(g_i)}{\sum_{i \in G} length(g_i)}$
    \item Weighted F1-Score: $wF1 = 2 \cdot \frac{wPrecision \cdot wRecall}{wPrecision + wRecall}$
\end{itemize}


\subsection{Quantitative Results}
Our adapted Smart-Tree algorithm demonstrates improvements over the original version across multiple grapevines, as shown in Table \ref{tab:performance_comparison}. The adapted algorithm achieved higher scores in unweighted and weighted metrics, indicating better overall performance and improved accuracy in detecting and reconstructing grapevine structures. Notably, recall enhanced from 0.63 to 0.78, a 23.8\% increase, while the F1-Score rose from 0.57 to 0.66, a 15.8\% improvement. Weighted metrics showed even more substantial enhancements, with weighted precision, recall, and F1-Score improving by 21.7\%, 23.8\%, and 22.9\%, respectively. The weighted IoU score increased from 0.59 to 0.73, representing a 23.72\% improvement in capturing the overall vine structure, particularly for longer, more significant branches. Additionally, the number of correctly identified canes increased from 17.85 to 21.42, while undetected canes decreased from 11.18 to 7.62, indicating better coverage of the grapevine structure.

\begin{table}[htbp]
\centering
\caption{Performance Comparison of Original and Adapted Smart-Tree}
\label{tab:performance_comparison}
\begin{tabular}{lcc}
\hline
Metric & Original Smart-Tree & Adapted Smart-Tree \\
\hline
Precision & 0.53 & 0.61 \\
Recall & 0.62 & 0.74 \\
F1-Score & 0.57 & 0.66 \\
Weighted Precision & 0.60 & 0.73 \\
Weighted Recall & 0.63 & 0.78 \\
Weighted F1-Score & 0.61 & 0.75 \\
Weighted IoU Score & 0.59 & 0.73 \\
Correct Canes & 17.85 & 21.42 \\
Undetected Canes & 11.18 & 7.62 \\
Erroneous Canes & 15.75 & 14.24 \\
\hline
\end{tabular}
\end{table}

\subsection{Qualitative Results}
Fig. \ref{fig:qualitative} shows the input point cloud, segmented point cloud, and final output model using our adapted Smart-Tree algorithm. The results demonstrate accurate structural alignment with the underlying grapevine architecture. A video demonstration of our method applied to a single vine can be viewed at \url{https://uc-vision.github.io/grapevine-3d-scanner-landing-page/}.








\subsection{Challenges and Limitations}
Despite improvements in grapevine skeletonization, several challenges persist. Segmentation errors in the initial point cloud can lead to broken or missed canes, which are difficult to rectify. Our algorithm sometimes incorrectly connects canes from neighbouring vines, distorting the vine structure. Additionally, gaps in point cloud reconstruction contribute to incomplete skeletal representations. The current heuristic used in the Minimum Spanning Tree stage sometimes fails to merge segments belonging to the same cane, resulting in relatively low precision. The validation of node detection accuracy presents another challenge, as it requires instance-level annotations of all nodes in the vine - data which is currently unavailable in our dataset.

\section{Conclusion}
This paper adapts the Smart-Tree algorithm for 3D grapevine modelling and introduces a graph-based approach for disambiguating skeletonization. Our method shows significant improvements, with an 15.8\% increase in F1-Score and 23.72\% in weighted IoU score. This enhanced model provides a foundation for more precise viticulture analysis and management. Future work will focus on developing an end-to-end deep learning approach for extracting skeletons from point clouds, and exploring applications in automated pruning and yield estimation, contributing to more informed decision-making in precision viticulture.

\section*{Acknowledgment}
The research reported in this article was conducted as part of “Predicting the
unseen: a new method for accurate yield estimation in viticulture/horticulture”,
which is funded by the New Zealand Ministry of Business, Innovation and
Employment (UOCX2308).

\bibliographystyle{IEEEtran}
\bibliography{ref}

\end{document}